\documentclass{PoS}

\title{Hadronic fluctuations and correlations at nonzero chemical potential}

\ShortTitle{Hadronic fluctuations and correlations}

\author{\speaker{Christian Schmidt}\\
        Universit\"at Bielefeld, Fakult\"at f\"ur Physik, 
	Universit\"atsstrasse 25, D-33615 Bielefeld, Germany\\
        E-mail: \email{schmidt@physik.uni-bielefeld.de}}

\abstract{We present a lattice study of fluctuations and correlations
among the conserved charges baryon number and strangeness 
in (2+1)-flavor QCD. The lattice calculations are based on a
Taylor expansion of the pressure. Results are presented at zero and
nonzero density on lattices with four and six time slices,
corresponding to a lattice spacing of $a\approx 0.25$~fm and
$a\approx 0.16$~fm, respectively. The quark masses are almost
physical, i.e.  the light quark mass has been chosen to one tenth of the
physical strange quark mass while the strange quark mass has been
set to its physical value. We find that all analyzed fluctuations and
correlations develop a peak at nonzero density, when they are treated
in an expansion up to sixth order. Especially below the critical
temperature ($T_c$) fluctuations and correlations increase
drastically, whereas above $T_c$ they are rather unaffected.}

\FullConference{5th International Workshop on Critical Point and Onset of Deconfinement - CPOD 2009,\\
		 June 08 - 12 2009\\
		 Brookhaven National Laboratory, Long Island, New York, USA}

\def \beq{\begin{equation}}
\def \eeq{\end{equation}}
\newcommand{\ave}[1]{\left\langle #1 \right\rangle} %regular average
\def\lsim{\raise0.3ex\hbox{$<$\kern-0.75em\raise-1.1ex\hbox{$\sim$}}}
\def\gsim{\raise0.3ex\hbox{$>$\kern-0.75em\raise-1.1ex\hbox{$\sim$}}}
\begin{document}

\section{Introduction}
Fluctuations of conserved charges, like baryon number, electric charge
and strangeness are generally considered to be sensitive indicators
for the structure of the thermal medium that is produced in heavy ion
collisions \cite{Koch}. In fact, if at non-vanishing baryon number a
critical point exists in the QCD phase diagram, this will be signaled
by divergent fluctuations of e.g. the baryon number density
\cite{Stephanov:1998dy}.

We present here results from lattice calculations of baryon number and
strangeness fluctuations in QCD with dynamical light and strange quark
degrees of freedom.  The results are based on calculations with an
improved staggered fermion action (p4-action) that strongly reduces
lattice cut-off effects in bulk thermodynamics at high
temperature. The values of the quark masses used in this calculation
are almost physical; the strange quark mass, $m_s$, is fixed to its
physical value while the light up and down quark masses are taken to
be degenerate and equal to $m_s/10$. This is about twice as large as
the average up and down quark masses realized in nature. We obtained
results from calculations performed with two different values of the
lattice cut-off, corresponding to lattices with temporal extent
$N_\tau=4$ and $6$. This allows us to judge the magnitude of
systematic effects arising from discretization errors in our improved
action calculations.  The spatial volume has been chosen to be
$V^{1/3}T=4$, which insures that finite volume effects are small.
 
At the QCD critical endpoint (CEP) the correlation
length of the chiral critical mode $\sigma$ will diverge. Correspondingly, 
all kinds of fluctuations of conserved charges, which couple to the $\sigma$-field,
become large in the vicinity of, and diverge at the CEP. It has been 
argued in \cite{Stephanov:1998dy,Stephanov:1999zu} that quadratic variances 
of event-by-event observables such as particle abundances, particle ratios 
or mean transverse momenta will reflect these divergent 
fluctuations and are thus good experimental observables for the
determination of the CEP.  It will thus be very interesting to study
baryon number and strangeness fluctuations as well as baryon number-strangeness 
correlations at nonzero density as they are related
to the event-by-event fluctuations of the proton, the kaon and their ratio. 
The latter has been recently measured by the NA49 collaboration for
central Pb-Pb collisions at five different SPS energies \cite{Stock}.

\section{The Taylor expansion method}

\begin{figure}
\begin{center}
\resizebox{0.32\textwidth}{!}{%
  \includegraphics{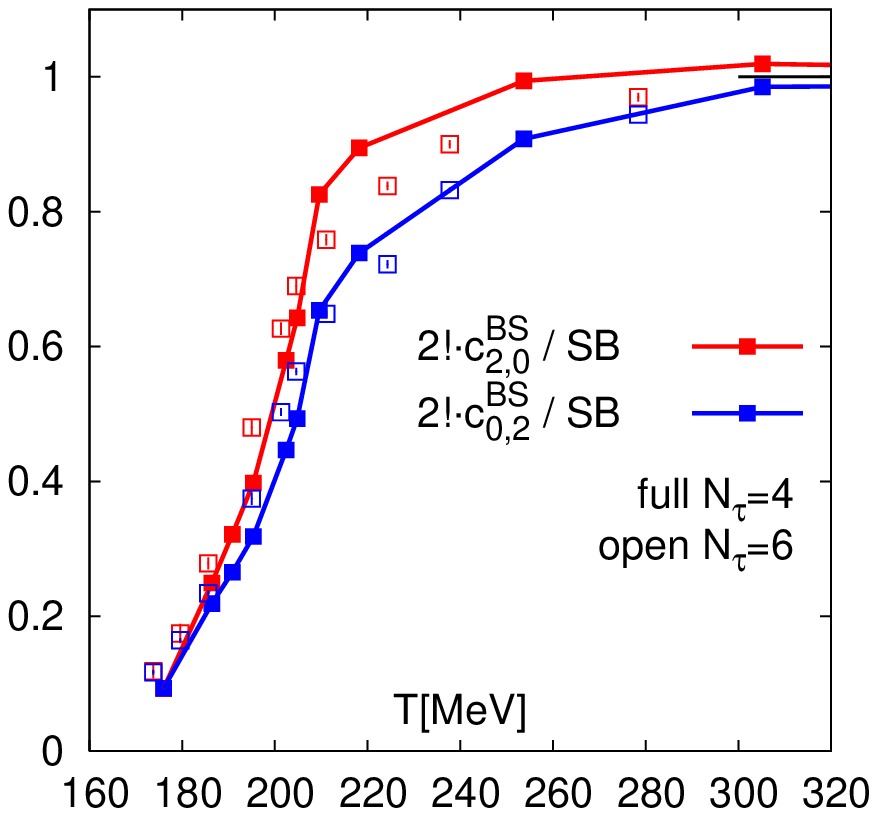}
}
\resizebox{0.32\textwidth}{!}{%
  \includegraphics{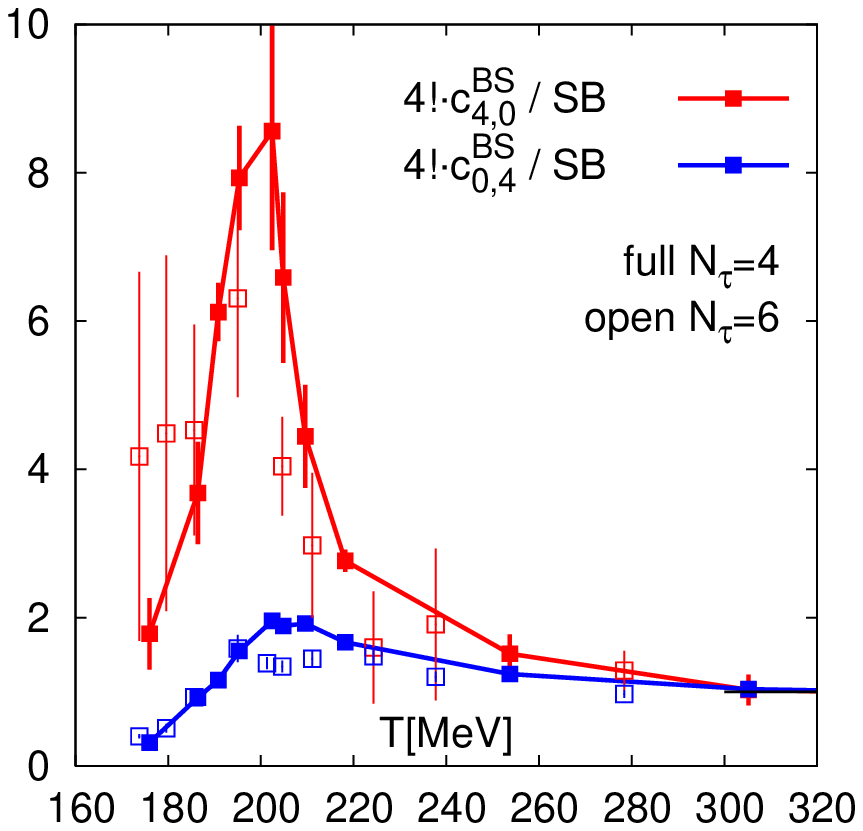}
}
\resizebox{0.32\textwidth}{!}{%
  \includegraphics{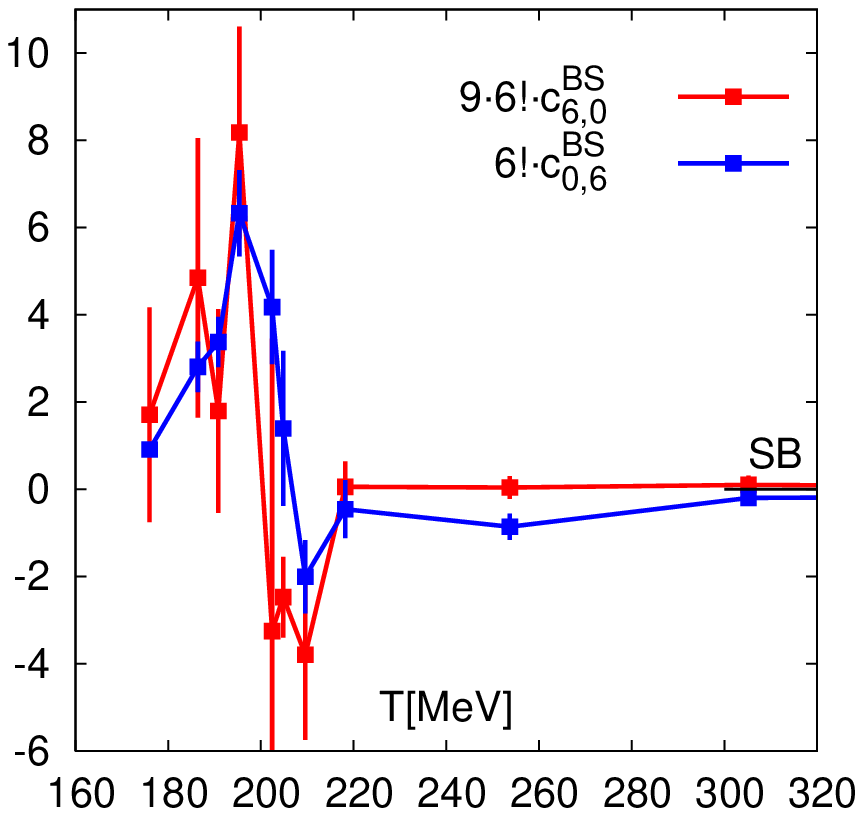}
}\\
\resizebox{0.32\textwidth}{!}{%
  \includegraphics{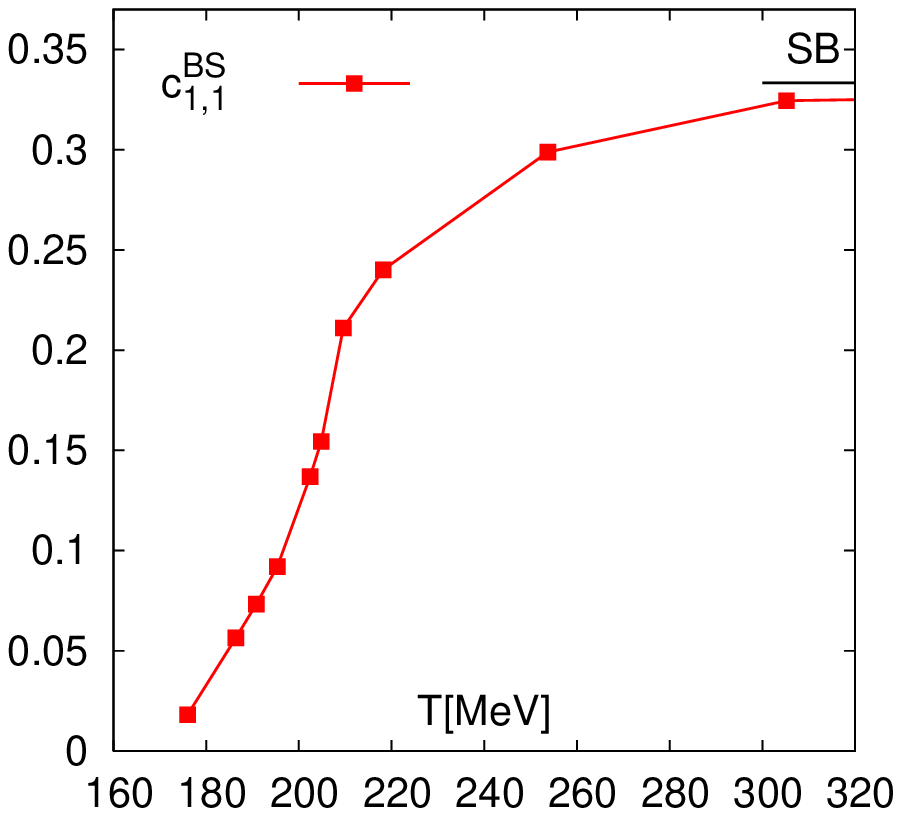}
}
\resizebox{0.32\textwidth}{!}{%
  \includegraphics{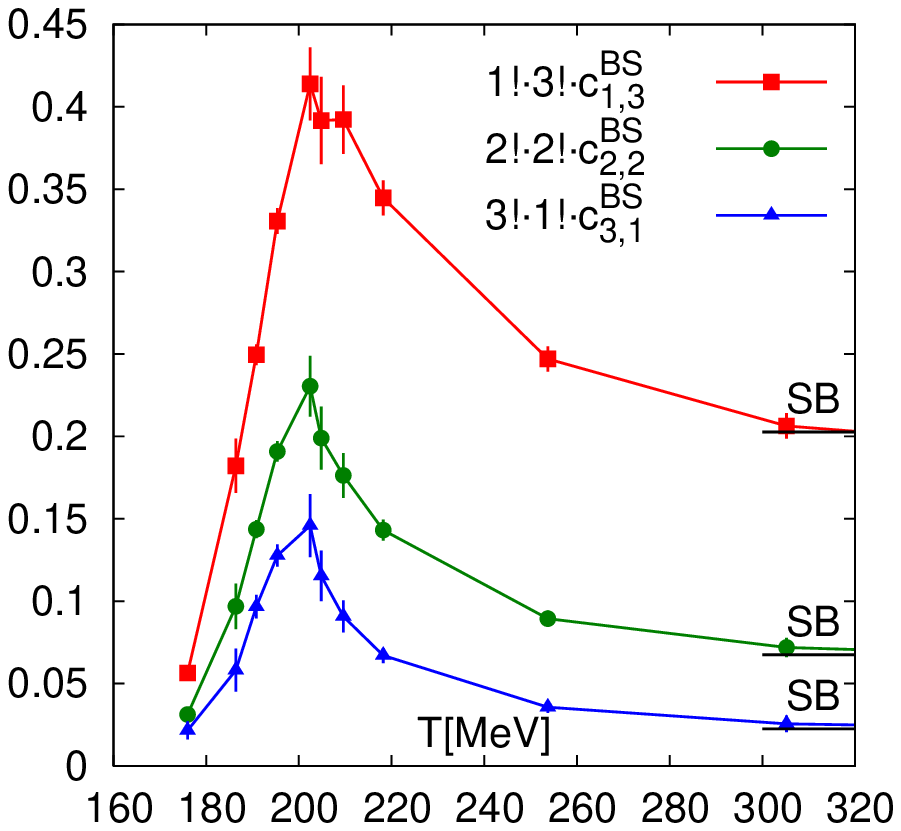}
}
\resizebox{0.32\textwidth}{!}{%
  \includegraphics{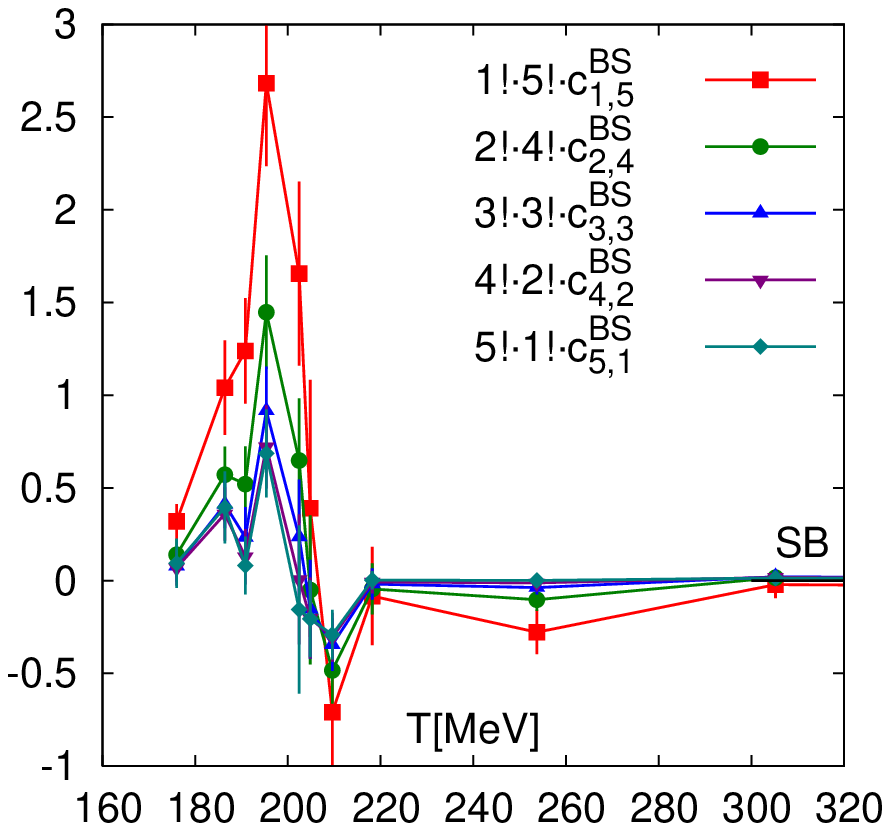}
}
\caption{Diagonal (upper row) and off-diagonal (lower row) expansion
coefficients of the pressure of 2nd (left column), 4th (middle column)
and 6th (right column) order. For diagonal coefficients up to the 4th
order we compare results from two different lattice spacings ($N_\tau = 4$
and $N_\tau =6$ lattices). Here we have normalized the data by its
corresponding SB-values.}
\label{fig:c_ij}
\end{center}
\end{figure}

Direct lattice calculations at nonzero baryon density are not possible
by means of standard Monte Carlo methods.  We follow here the Taylor
expansion approach to finite density QCD on the lattice, as described
in detail in \cite{our, Allton:2002zi}.  Starting from an expansion
of the logarithm of the QCD partition function, {\it i.e.} the
pressure, one obtains
\beq
\frac{p}{T^4} \equiv \frac{1}{VT^3}\ln Z(V,T,\mu_B,\mu_S)  
=\sum_{i,j}c_{i,j}^{B,S}\left(\frac{\mu_B}{T}\right)^i
\left(\frac{\mu_S}{T}\right)^j\; ,
\label{pressure}
\eeq
where
\beq
c_{i,j}^{B,S}(T)=\frac{1}{i!j!}\left.\frac{\partial^i}{\partial \mu_B^i}
\frac{\partial^j}{\partial \mu_S^j}
\frac{\ln Z(V,T,\mu_B,\mu_S) }{VT^3}\right|_{\mu_B=\mu_S=0}\; .
\label{coeffBS}
\eeq
Here $\mu_B$ and $\mu_S$ are the baryon and strangeness chemical
potentials, respectively, which can be obtained as appropriate linear combinations
of the quark chemical potentials. Note that we treat the up and down quarks as
degenerated, both in mass as well as net quark number
density. Accordingly, isospin and electric charge chemical potential
vanish. Due to charge conjugation symmetry only coefficients with
$i+j$ even are nonzero. In Fig.~\ref{fig:c_ij} we show the
diagonal and off-diagonal coefficients of this expansion up to 6th order. 
For diagonal coefficients up to 4th order we show results that
have been obtained for different lattice spacings, i.e. on $N_\tau =4$ 
and $N_\tau =6$ lattices. 
We note that results obtained on lattices with temporal extent
$N_\tau =6$ are in good agreement with those obtained on the 
coarser $N_\tau=4$ lattice. The slight shift towards smaller 
temperatures, visible for the $N_\tau=6$ data
relative to the $N_\tau=4$ data, is consistent with findings for 
the equation of state, e.g. the trace anomaly $(\epsilon -3p)/T^4$,
and also reflects the shift in the transition temperature
observed when comparing the locations of the cusp in the chiral 
susceptibility \cite{our_Tc}.

\begin{figure}
\begin{center}
\resizebox{0.48\textwidth}{!}{%
  \includegraphics{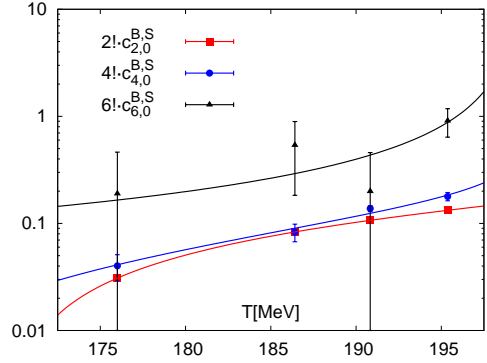}
}
\caption{A fit to the expansion coefficients of the pressure below
$T_c$, inspired by the critical behavior of the free energy. At this
lattice size and quark mass ($N_\tau=4$, $m_q=0.1m_s$) we have
$T_c\approx$ 202 MeV.}
\label{fig:fit}
\end{center}
\end{figure}

The general pattern of the coefficients can be understood from the 
structure of the singular part of the free energy. In the chiral limit
of our (2+1)-flavor simulations, {\it i.e.} with vanishing light quark
masses, but finite and physical strange quark mass, we expect the 
QCD transition to be of second order \cite{our_scaling}. The relevant
scaling field which contains the baryon chemical potential will be
the reduced temperature, for which we make the ansatz
\beq
t=\frac{T-T_c}{T_c}+\kappa\left(\frac{\mu_B}{T}\right)^2\;.
\eeq
Accordingly two derivatives of the free energy with respect to $\mu_B$
(at $\mu_B=0$) are similar to one derivative with respect to the
temperature. Therefore we obtain the following formulas, which can be
used to fit the critical behavior of the Taylor expansion coefficients
\begin{eqnarray}
2!\cdot c_{B,S}^{2,0}&\sim& 
\mp 2 A_{\pm} (2-\alpha) \kappa |t|^{1-\alpha}+b_2 t+ c_2 \; ;
\label{scaling_c2}\\
4!\cdot c_{B,S}^{4,0}&\sim& 
-12 A_{\pm} (2-\alpha)(1-\alpha) \kappa^2 |t|^{-\alpha}+b_4 t+ c_4 \; ; 
\label{scaling_c4}\\
6!\cdot c_{B,S}^{6,0}&\sim& 
\pm 120 A_{\pm} (2-\alpha)(1-\alpha)(-\alpha) \kappa^3 |t|^{-1-\alpha} \; .
\label{scaling_c6}
\end{eqnarray}
Here uppers signs are valid for $t>0$, whereas lower signs are for $t<0$. 
We see that the critical behavior is governed by the critical
exponent $\alpha$, which is small and negative for the universality classes 
of interest. For staggered fermions
at finite lattice spacing, were the chiral symmetry is broken down to
a $U(1)$ subgroup, we expect the relevant universality class to be that of
the 3d-$O(2)$ symmetric model, where we have $\alpha\sim -0.015$. We
thus expect $c_{B,S}^{4,0}$ to develop a cusp, while $c_{B,S}^{6,0}$
will diverge in the chiral limit. $c_{B,S}^{2,0}$ will be dominated by
the regular part of the free energy. Hence, we took into
account the leading order regular terms, indicated by the coefficients
$b_2, b_4, c_2$, and $c_4$ in Eq.~\ref{scaling_c2} and Eq.~\ref{scaling_c4}. For
$T<T_c$ we show the fit results in Fig.~\ref{fig:fit}.
Although the ansatz is strictly valid only in the chiral limit, the fit
works quite well. It has already been observed previously that at this
quark mass ($m_q = 0.1m_s$) the QCD thermodynamics enters the scaling
regime of the critical behavior \cite{our_scaling}.  Currently the coefficient $\kappa$,
which determines the slope of the critical line as function of $\mu_B$
is, however, not very well constrained.

\section{Baryon number and strangeness fluctuations}
\begin{figure}
\begin{center}
\resizebox{0.32\textwidth}{!}{%
  \includegraphics{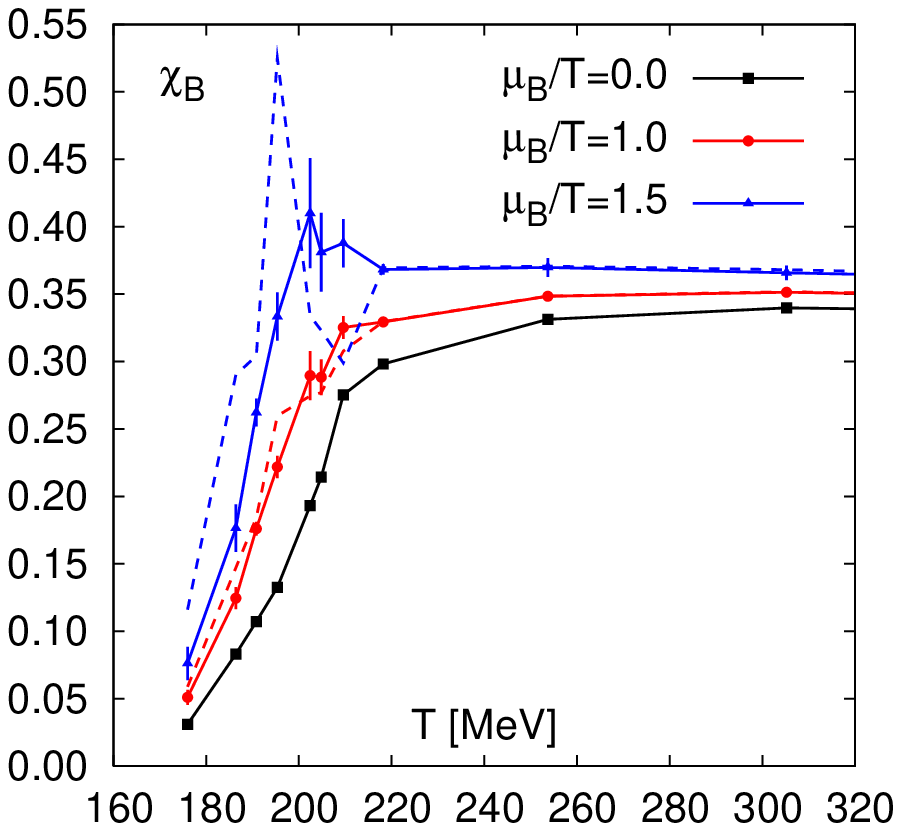}
}
\resizebox{0.32\textwidth}{!}{%
  \includegraphics{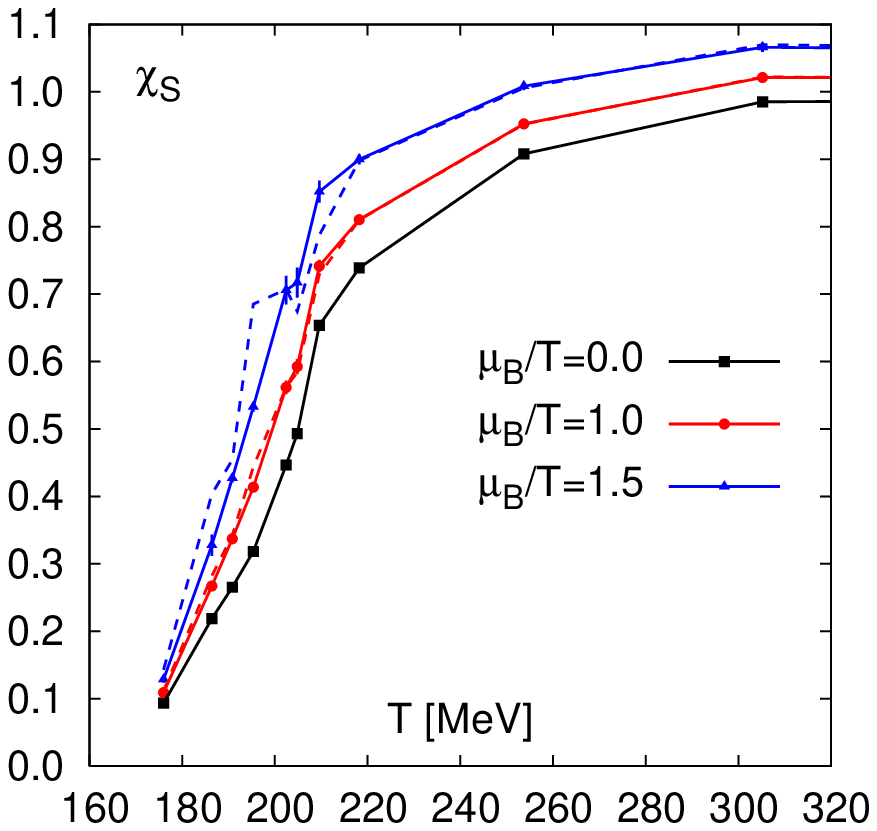}
}
\resizebox{0.32\textwidth}{!}{%
  \includegraphics{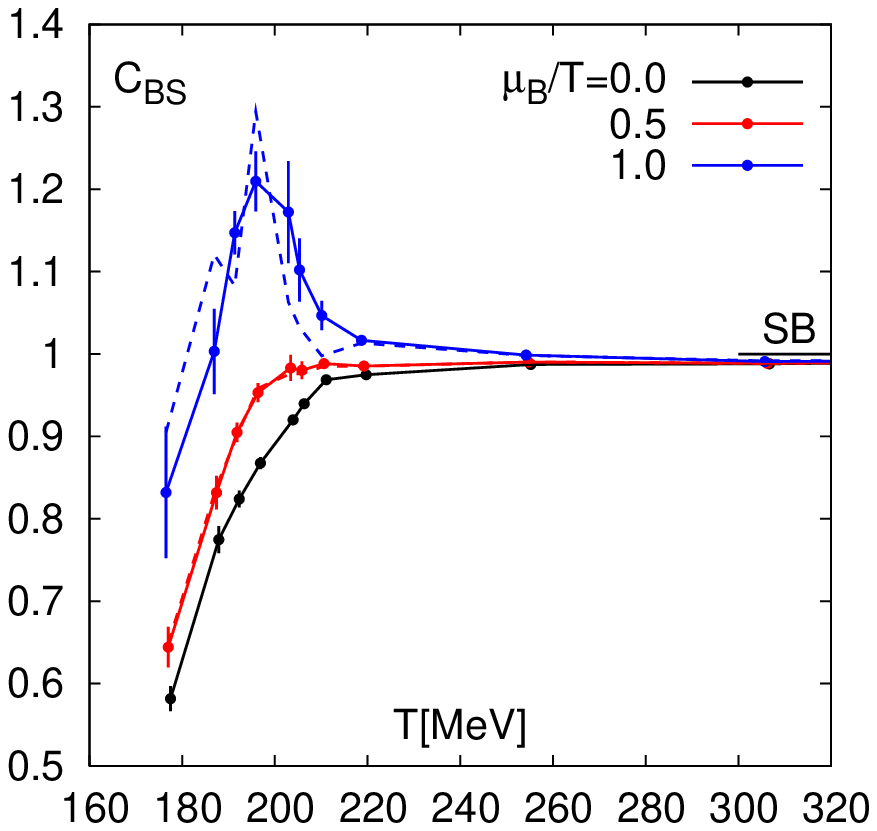}
}\\
\caption{Baryon number fluctuations (left), strangeness fluctuations
(middle) and baryon number strangeness correlation coefficient (right)
at nonzero baryon chemical potential. Data points are obtained by a
truncated Taylor series up to the 4th order in $\mu_B/T$, while dashed
lines indicate the 6th order results.}
\label{fig:fluct}
\end{center}
\end{figure}

We now construct baryon number and strangeness susceptibilities which measure 
fluctuations at nonzero
chemical potentials. The corresponding expansions can be expressed in
terms of the pressure coefficients from Eq.~\ref{pressure}. We obtain
for the baryon number fluctuations
\beq
\chi_B\equiv\frac{\left<N_B^2\right>-\left<N_B\right>^2}{VT^3}
=\sum_{i=1}^{\infty}(2i)(2i-1)c_{B,S}^{2i,0}\left(\frac{\mu_B}{T}\right)^{2i-2}\; ,
\label{series_B}
\eeq
and for the strangeness fluctuations
\beq
\chi_S\equiv\frac{\left<N_S^2\right>-\left<N_S\right>^2}{VT^3}
=\sum_{i=0}^{\infty}2c_{B,S}^{2i,2}\left(\frac{\mu_B}{T}\right)^{2i}\; .
\label{series_S}
\eeq
Here $\ave{\cdot}$ indicates the expectation value evaluated in the grand 
canonical ensemble in
lattice simulations. $N_B$ and $N_S$ are the net number of baryons and strange
particles, respectively. In Fig.~\ref{fig:fluct} (left and middle) we 
show results for the baryon number and strangeness fluctuations
at zero and nonzero $\mu_B/T$, obtained by truncating the series 
(\ref{series_B} and \ref{series_S}) after the 4th order.
We have also indicated the 6th order results by dashed lines in order to
give a feeling for the truncation errors. At $\mu_B/T=0$, the generic form of the temperature
dependence is a smooth crossover, while the leading order term in $(\mu_B/T)^2$
has a peak at $T_c$. It is thus clear that at sufficiently large values of
$\mu_B/T$ the fluctuations will develop a peak. However, it is important to 
stay within the radius of convergence of the series, in order to keep the 
truncation errors small. We will discuss the radius of convergence in 
Sec.~\ref{sec:radius} in more detail; here we find that for $\mu_B/T\leq 1.5$
the truncation error is small. For this value of the chemical potential the
baryon number fluctuations show already a small peak, whereas the strangeness
fluctuations are only slightly enhanced.

We note, that all fluctuations of modes that couple to the relevant chiral 
critical mode at the QCD critical point (the sigma-field) will diverge at
the QCD critical endpoint (CEP). As light quarks carry also a baryon number,
the baryon number fluctuations are expected to diverge at the CEP.

\section{The baryon number  -- strangeness correlation}
The baryon number-strangeness coefficient was introduced as a diagnostic 
tool for the effective degrees of freedom in the quark gluon plasma
\cite{Koch:2005vg}. It is defined as
\beq
C_{BS} \equiv 
-3 \frac{\ave{N_B N_S}-\ave{N_B}\ave{N_S}}{\ave{N_S^2} - \ave{N_S}^2}\; .
\eeq
 Lattice calculations of this quantity in the
(partially) quenched approximation \cite{Gavai:2005yk} and
$2+1$-flavor QCD \cite{our} indicate that this quantity reaches unity
for temperatures larger than $1.5T_c$, whereas it is significantly
smaller for $T<T_c$. {\it I.e.}, the quark flavors are uncorrelated
above $T_c$ as in the ideal Quark Gluon Plasma (QGP). This behavior
indicates the transition from hadronic degrees of freedom to that of a
QGP.

Similar to the expansion of baryon number fluctuations (Eq. \ref{series_B}) and 
strangeness fluctuations (Eq. \ref{series_S}) we can also expand the baryon number-
strangeness coefficient. Again we express its expansion coefficients in terms of 
the expansion coefficients of the pressure (Eq.~\ref{pressure}) and obtain
\beq
C_{BS}(\mu_B/T)
=C_{BS}^{(0)}
+C_{BS}^{(2)}\left(\frac{\mu_B}{T}\right)^2
+C_{BS}^{(4)}\left(\frac{\mu_B}{T}\right)^4
+\mathcal{O}\left[\left(\frac{\mu_B}{T}\right)^6\right]\;,
\eeq
with
\beq
C_{BS}^{(0)}=-\frac{3c_{1,1}^{B,S}}{2c_{0,2}^{B,S}}\; ; \quad
C_{BS}^{(2)}=\frac{3c_{1,1}^{B,S} c_{2,2}^{B,S} 
- 9 c_{0,2}^{B,s} c_{3,1}^{B,S}}{2\left(c_{0,2}^{B,S}\right)^2}\; ;
\eeq
and
\beq
C_{BS}^{(4)}=\frac{-3 c_{1,1}^{B,S} \left(c_{2,2}^{B,S}\right)^2 
+9 c_{0,2}^{B,S} c_{2,2}^{B,S} c_{3,1}^{B,S}
+3 c_{0,2}^{B,S} c_{1,1}^{B,S} c_{4,2}^{B,S}
-15 \left(c_{0,2}^{B,S}\right)^2 c_{5,1}^{B,S} 
         }{2\left(c_{0,2}^{B,S}\right)^3}\;.
\eeq

In Fig.~\ref{fig:fluct} (right) we show the correlation $C_{BS}$ at 
zero and nonzero density. 
We find that at nonzero baryon number density
a peak is developing already at small values of $\mu_B/T$ where the
truncation errors are small (data points are the 4th order results, while
the dashed lines indicate the 6th order results).
At $\mu_B/T=1.0$, the correlations at
$T_c$ increase by almost 50\%, and rise to a value which is about 20\%
larger than the Stefan-Boltzmann value. It will be interesting to see whether the 
increase in the baryon number strangeness coefficient will give rise to a peak in 
proton over kaon fluctuations obtained in Heavy Ion Collisions 
and whether these fluctuations can be used as an experimental signal
for locating the critical point.
Also below $T_c$ the correlations increase drastically with increasing
$\mu_B$, whereas above $T_c$ the correlation coefficient $C_{BS}$ remains rather 
unaffected by the increasing baryon density.

\section{The radius of convergence}
\label{sec:radius}
It is known that in the temperature region close to $T_c$ the convergence
properties of the Taylor expansion of the pressure (Eq.~\ref{pressure}) 
and therefore also of the expansions of the baryon number and strangeness 
fluctuations (Eq.~\ref{series_B}) and (Eq.~\ref{series_S}) are poor. 
Here the expansion coefficients wildly fluctuate in sign and magnitude. 
Only for temperatures below the CEP, this might be better as in this case 
all expansion coefficients are positive. It has 
been argued that the radius of convergence can be used to determine the 
location of the CEP \cite{radius}. One way to define the radius of convergence of the 
pressure series in $\mu_B/T$ (Eq.~\ref{pressure}) is
\beq
\rho=\lim_{n\to\infty}\rho_n
\quad\mbox{with}\quad
\rho_n=\mu_B^{(n)}/T^{(n)}=\sqrt{c_{n,0}^{B,S}/c_{n+2,0}^{B,S}}\; .
\label{rho}
\eeq
The convergence radius can be estimated in a similar manner by the coefficients 
in the series of the baryon 
number fluctuations (Eq.~\ref{series_B}). 
Each order $\rho_n$ will, however, differ by a constant factor which
goes to one in the limit $n\to\infty$. We define a second estimator of the 
convergence radius, based on the $\chi_B$-series 
as
\begin{equation}
\rho_n[\chi_B]=\sqrt{\frac{n(n-1)}{(n+1)(n+2)}}\;\rho_n[p/T^4]\;.
\label{rhochiB}
\end{equation}
The method to estimate the location of the CEP by the radius of convergence
works in two steps:
\begin{enumerate}
\item Find the largest temperature where all (available) expansion coefficients are positive. 
Only if all expansion coefficients are positive, the corresponding singularity 
which is limiting the convergence radius lies on the real axis and can be associated
with the physical phase transition.
\item Estimate the radius of convergence at this temperature by using Eq.~\ref{rho} or Eq.~\ref{rhochiB}.
\end{enumerate}
It is clear, that in practice we can not perform the limit $n\to\infty$ but have to 
stop at some finite $n$. Eventually at this point different estimators of the radius of 
convergence will agree within errors.

\begin{figure}
\begin{center}
\resizebox{0.48\textwidth}{!}{%
  \includegraphics{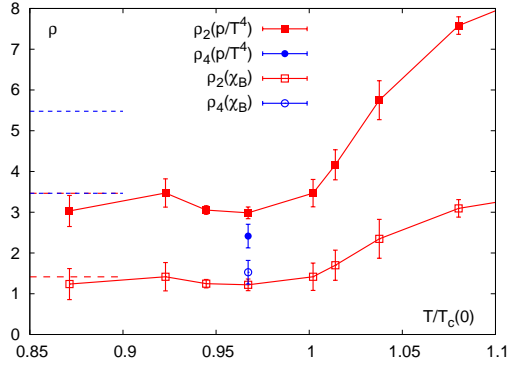}
}
\caption{Two estimates of the radius of convergence of the pressure series (full symbols) and 
the series of baryon number fluctuations (open symbols) as a function of temperature. $\rho_4$ 
indicates our best estimate for the location of the critical endpoint. Dashed lines show the 
resonance gas limit of the different estimators.}
\label{fig:rho}
\end{center}
\end{figure}

In Fig.~\ref{fig:rho} we have summarized current results on the radius of convergence.
The first non trivial approximation which does not explicitly include contributions from 
the gluon sector (contributes to $c_{0,0}^{B,S}$), is given by $\rho_2$, shown by the square data
points. Here both contributing expansion coefficients ($c_{2,0}^{B,S}$ and $c_{4,0}^{B,S}$) are positive 
for all values of the temperature,
hence an approximation for the temperature ($T^{(2)}$) of the CEP can not be deduced. At this
order the estimates from the pressure series (full symbols) and from the series of the baryon 
number fluctuations differ quite dramatically. The next higher 
approximation $\rho_4$ of the CEP, which is currently our best approximation, is shown by the 
circle data point. We obtain for the temperature $T^{(4)}\approx 0.96 T_c(\mu_B=0)$. 
Here the difference between the two estimators becomes less severe. For the convergence radius they 
span the range of $\rho\approx(1.5-2.4)$. It is interesting to note that the two different estimators 
of the convergence radius seem to approach the $n\to\infty$ limit from different sites. 
However, the $\chi_B$-estimator seems to be the more stable one.
Also shown are the resonance gas limits of the two approximations, indicated
by the dashed lines. 

\section{Taylor expansion vs. Pad\'{e} approximation}
\begin{figure}
\begin{center}
\resizebox{0.48\textwidth}{!}{%
  \includegraphics{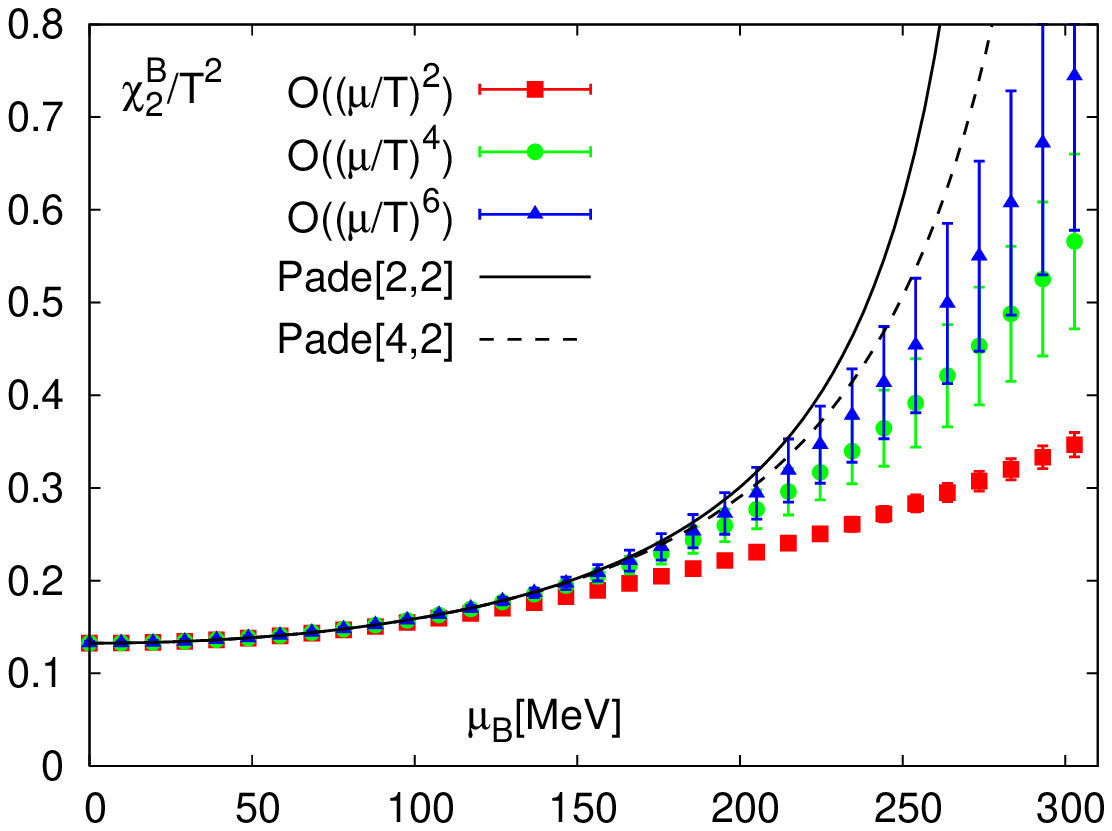}
}
\resizebox{0.48\textwidth}{!}{%
  \includegraphics{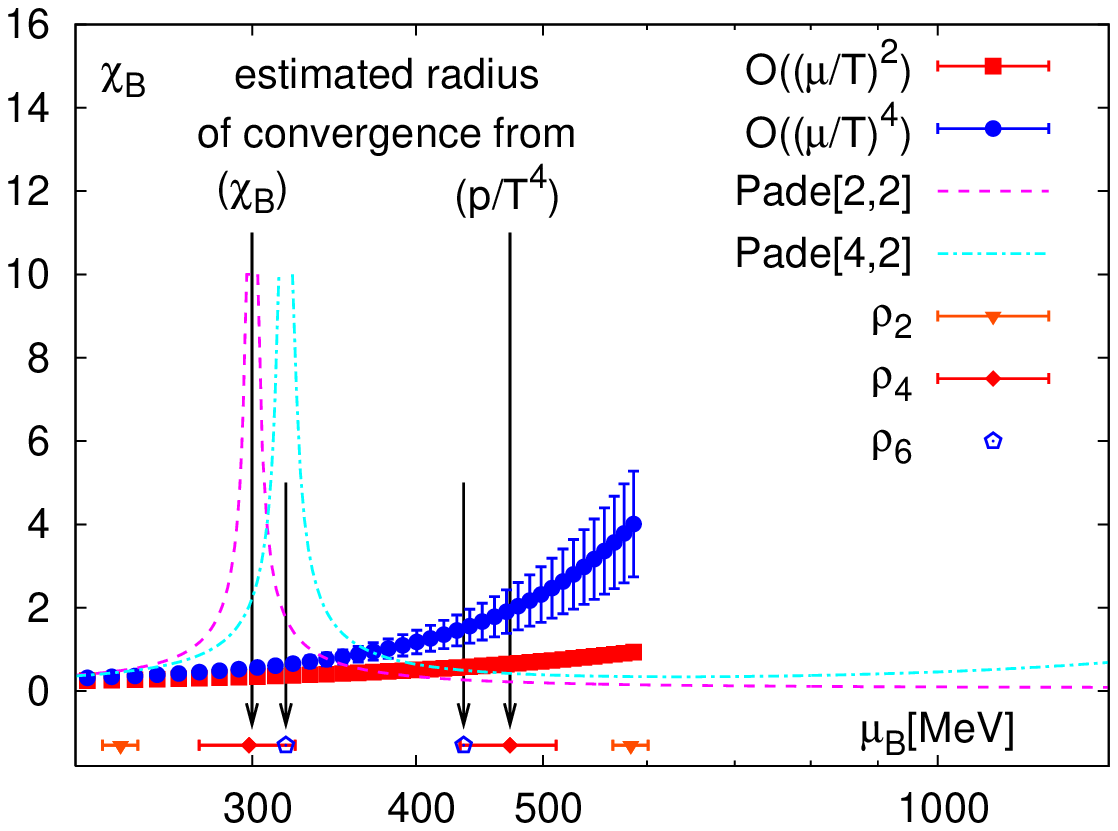}
}
\caption{Baryon number fluctuations at $T=0.96 T_c \approx T^{\rm CEP}$ as a
function of the baryon chemical potential (left). We compare different
order of the truncated Taylor series vs two Pad\'{e} approximants. The right panel
shows the same but on a much larger scale.}
\label{fig:pade}
\end{center}
\end{figure}

So far our analysis of observables at nonzero baryon density has been based on the Taylor 
expansion of the pressure (Eq.~\ref{pressure}) around $\mu_B/T=0$. Now we want to discuss the 
Pad\'{e} approximation, which is usually considered to be a better approximation of the true function,
even beyond the radius of convergence \cite{Pade}. One can construct different Pad\'{e} approximants from the 
Taylor expansion coefficients. Pad\'{e} 
approximants are rational functions which differ by the order 
of the polynomials in the numerator and denominator. For the baryon number fluctuations we obtain e.g.
\beq
{\rm Pade}[2,2]\left(\chi_B\right)=\frac{4 c_{B,S}^{2,0} c_{B,S}^{4,0} + 2 \left(12 \left(c_{B,S}^{4,0}\right)^2 
- 5 c_{B,S}^{2,0} c_{B,S}^{6,0}\right)\left(\frac{\mu_B}{T}\right)^2}
{2 c_{B,S}^{4,0} - 5 c_{B,S}^{6,0} \left(\frac{\mu_B}{T}\right)^2}
\label{pade22}
\eeq
In Fig.~\ref{fig:pade}(left) we compare different orders of the truncated Taylor series (Eq.~\ref{series_B})
with the Pad\'{e} approximations [2,2] and [4,2], where the latter one has a fourth order polynomial in 
the numerator. 
We show the baryon number fluctuations at fixed, but
varying baryon chemical potential. 
$T\approx 0.96 T_c$ is 
currently our best approximation for the temperature of the critical
endpoint ($T^{\rm CEP}$) as it was estimated by the radius of
convergence method. The most prominent feature of the Pad\'{e}
approximations is the appearance of a pole. The location of the 
poles are given by the roots of the denominator. From Eq.~\ref{pade22}
it is obvious that the pole in the [2,2] approximation exactly coincides
with the $\rho_4[\chi_B]$ estimate of the convergence radius. 

Note that for the Taylor series of $\chi_B$ which was truncated after the 6th
order, as well as for the [4,2] Pad\'{e} approximant we need the
coefficient $c_{B,S}^{8,0}$. In the following analysis we set the so far unknown 
coefficient to $c_{B,S}^{8,0}\equiv c_{B,S}^{6,0}/x$. 

Given the coefficients $c_{B,S}^{2,0}$ -- $c_{B,S}^{6,0}$ at $T\approx 0.96 T_c$, 
we vary the strength of $c_{B,S}^{8,0}$ by the parameter $x$. We have fixed the 
parameter $x$ by assuming that the asymptotic behavior of the [4,2] Pad\'{e} 
approximant for large $\mu_B$ should be governed by the free gas result which 
is given by
\begin{equation}
\lim_{\mu_B\to\infty}\chi_B\approx \frac{1}{3} 
+ \frac{1}{9\pi^2}\left(\frac{\mu_B}{T}\right)^2\;.
\end{equation}
Using this property, we obtain $x\approx 4.95$ and the next approximation for 
the convergence radius ($\rho_6=\sqrt{x}$), which we also shown in 
Fig.~\ref{fig:pade}(right). Again we find that the Pole in the [4,2] Pad\'{e} 
approximant for $\chi_B$ coincides exactly with the corresponding estimator 
for the convergence radius $\rho_6[\chi_B]$.

\section{Summary and conclusions}
We have performed a comprehensive study of the Taylor expansion coefficients
of the baryon number fluctuation, strangeness fluctuation as well as their
correlations, based on the Taylor expansion coefficients of the pressure. 
We found that all these quantities will develop a peak at $T\approx T_c$ 
for nonzero baryon chemical potential, which is the dominant effect from the leading order
coefficient in $\mu_B$. The sub-leading coefficient will in general lead to
a shift of the peak, i.e. a $\mu_B$-dependence of the peak position. 
Our current analysis takes into account the coefficients up to 6th order. On the 
highest order the errors are barely under control. As most observables at nonzero 
chemical potential crucially depend on the relative strength of the coefficients, 
a more refined analysis which also takes into account the 8th order coefficients 
is highly desired. This will also help to confirm the 
structure of the Taylor expansion coefficients which can be understood in terms of an 
appropriate scaling ansatz of the free energy in the chiral limit. With a combined 
fit of all Taylor expansion coefficients it will in principle be possible to 
determine the $\mu_B$-dependent curvature of the critical temperature in the chiral limit.

When comparing different orders of the truncated Taylor series we find small
truncation errors below $\mu_B/T\lsim (1-1.5)$, while for $\mu_B/T\approx (1.5-2.5)$ 
consecutive
orders become compatible. The latter fact can be used to estimate the radius 
of convergence of the Taylor series, which will be connected with a physical 
singularity in the QCD phase diagram (critical endpoint) when all expansion 
coefficients are positive. The temperature of the critical endpoint we currently
approximate to $T^{\rm CEP}\approx 0.96T_c$.

We also constructed Pad\'{e} approximants from the Taylor coefficients. Again 
we find good agreement between different approximants as well as the truncated
Taylor series below $\mu_B/T\lsim (1-1.5)T_c$.

\section*{Acknowledgments}
\label{ackn}
This work has been supported in part by contracts DE-AC02-98CH10886
and DE-FG02-92ER40699 with the U.S. Department of Energy,
the Bundesministerium f\"ur Bildung und Forschung under grant
06BI401, the Gesellschaft
f\"ur Schwerionenforschung under grant BILAER and the Deutsche
Forschungsgemeinschaft under grant GRK 881. Numerical simulations have
been performed
on the QCDOC computer of the RIKEN-BNL research center, the DOE funded
QCDOC at BNL, the apeNEXT at Bielefeld University and the BlueGene/L
at the New York Center for Computational Sciences (NYCCS).

\end{document}